\documentclass[twocolumn,eqsecnum,aps,prb]{revtex4}% Physical Review B

\usepackage{graphicx}%
\usepackage{dcolumn}
\usepackage{amsmath}

\makeatletter
\def\btt#1{\texttt{\@backslashchar#1}}%
\DeclareRobustCommand\bblash{\btt{\@backslashchar}}%
\makeatother

\begin{document}

\title{Transport properties in CeOs$_{4}$Sb$_{12}$: Possibility of the ground state being semiconducting}

\author{H.~Sugawara,\cite{Tokushima}  S.~Osaki, M.~Kobayashi, T.~Namiki, S.R.~Saha, Y.~Aoki, and H.~Sato}

\affiliation{Department of Physics, Tokyo Metropolitan University, Minami-Ohsawa, Hachioji, Tokyo 192-0397, Japan}

\date{                  }

\begin{abstract}
We have measured both magnetoresistance and Hall effect in CeOs$_4$Sb$_{12}$ to clarify the large resistivity state ascribed to the Kondo insulating one and the origin of the phase transition near 0.9~K reported in the specific heat measurement. We found unusual temperature ($T$) dependence both in the electrical resistivity $\rho\sim T^{-1/2}$ and the Hall coefficient $R_{\rm H}\sim T^{\rm -1}$ over the wide temperature range of about two order of magnitude below $\sim30$~K, which can be explained as a combined effect of the temperature dependences of carrier density and carrier scattering by spin fluctuation. An anomaly related with the phase transition has been clearly observed in the transport properties, from which the $H-T$ phase diagram is determined up to 14~T. Taking into account the small entropy change, the phase transition is most probably the spin density wave one. Both the electrical resistivity and Hall resistivity at 0.3~K is largely suppressed about an order of magnitude by magnetic fields above $\sim3$~T, suggesting a drastic change of electronic structure and a suppression of spin fluctuations under magnetic fields.
\\
\end{abstract}


\maketitle
\section{INTRODUCTION}
\begin{figure}[b]
\begin{center}
\includegraphics[width=1\linewidth]{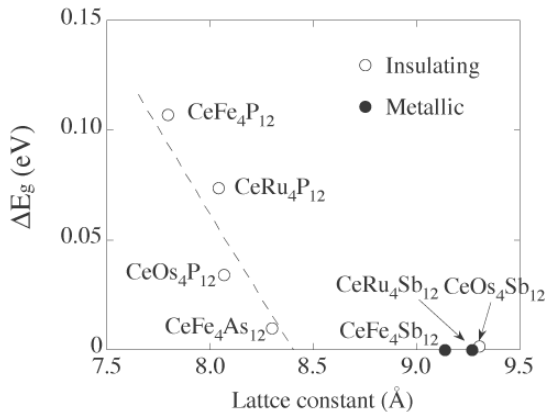}
\end{center}
\caption{Energy gap  $\Delta E_{\rm g}$ $vs$ lattice constant in CeTr$_{4}$Pn$_{12}$.}
\label{Eg}
\end{figure}

Filled-skutterudite compounds $RT_4$Pn$_{12}$ (R: rare earth, T: Fe, Ru, Os, and Pn: pnictogen),~\cite{Jeitschko} exhibit a wide variety of exotic phenomena associated with the unique body-centered cubic structure.~\cite{Sekine,Sato,Sugawara,Aoki,Bauer} Strong hybridization between 4$f$-electrons and conduction electrons enhanced by the large coordination number; 12 Pn and 8 $T$ ions surrounding $R$, is thought to realize such exotic phenomena.~\cite{Sugawara,Sato2} In fact, the energy gap $\Delta E_{\rm g}$ estimated from the temperature dependence of electrical resistivity $\rho(T)$ for the Ce-based filled skutterudites can be roughly scaled with the lattice constant as shown in Fig.~\ref{Eg}; the smaller lattice constant ones such as CeRu$_4$P$_{12}$ and CeFe$_4$P$_{12}$ have the larger energy gap.~\cite{Meisner,Shirotani,Grandjean,Morelli,Bauer2,Bauer3} 
%%%%%%%%%%%%%%%%%%%%
%%%%%%%%%%%%%%%%%%%%
Note the metal-insulator transition in PrRu$_{4}$P$_{12}$ and the apparent Kondo-like behaviors in PrFe$_{4}$P$_{12}$ also reflect strong $c-f$ hybridization,~\cite{Sekine,Sato,Aoki} which are unusual as Pr compounds.  In contrast, CeRu$_{4}$Sb$_{12}$ with a larger lattice constant is a metal exhibiting non-Fermi-liquid (NFL) behaviors at low temperatures,~\cite{Takeda,Bauer3} and PrRu$_{4}$Sb$_{12}$ is an ordinary superconductor exhibiting no exotic behaviors.

Thus, only from the lattice constant, PrOs$_{4}$Sb$_{12}$ was expected to be an ordinary metal, however, it was found to be the first Pr-based heavy fermion superconductor.~\cite{Bauer} CeOs$_{4}$Sb$_{12}$, predicted to be a metal by the band structure calculation,~\cite{Harima} was first reported to be a semiconductor with a small gap ($\Delta E_{\rm g}/{\bf k}_{\rm B}\sim10$~K) estimated from $\rho(T)$,~\cite{Bauer2} while the finite value of the Sommerfeld coefficient $\gamma \simeq0.2$J/K$^2$mol suggests metallic ground state.~\cite{Bauer2,Namiki} Recently, in a far-infrared measurement,~\cite{Matsunami} an apparent decrease of the optical conductivity at low temperatures, indicating the development of gap structure near the Fermi level $E_{\rm F}$ with decreasing temperature. The another controversial feature is the sharp peak observed in the specific heat $C(T)$ at $\sim1.1$~K in zero field, that was first ascribed to a phase transition of some impurity phase by Bauer {\it et al.}~\cite{Bauer2}  The field dependence of $C(T)$ reported recently by Namiki {\it et al}. rules out such a possibility of impurity effect and suggests the existence of intrinsic phase transition in this compound.~\cite{Namiki}  However, neither the origin of the peak nor the ground state of this material has been clarified at this stage. 
In this paper, we report the extended study of electrical resistivity ($\rho$) and Hall coefficient ($R_{\rm H}$) on high quality single crystals, to deepen the understanding of the ground state properties in CeOs$_{4}$Sb$_{12}$.

\section{EXPERIMENTAL}
High quality single crystals of CeOs$_{4}$Sb$_{12}$ and LaOs$_{4}$Sb$_{12}$ were grown by Sb-flux method starting from a composition of $R$:Os:Sb = 1:4:20 using raw materials 3N5 (99.95\%)- La, Ce, 3N-Os, and 6N-Sb.~\cite{Takeda,Bauer3} The lattice constants determined by x-ray powder diffraction agree with the reported values,~\cite{Braun} and absence of impurity phases was confirmed within the experimental accuracy. 
$\rho$ and $R_{\rm H}$ were measured by the ordinary dc four-probe method. The voltage measurements were made by Keithley 182 nanovoltmeters. In order to reduce the heating effect, samples were directly immersed in liquid $^3$He in the magnetoresistance (MR) and Hall resistivity ($\rho_{\rm H}$) measurements using an Oxford Instrument top loading $^3$He cryostat, down to 0.3~K and up to 14~T.  The magnetic measurements were made by a Quantum Design SQUID magnetometer up to 7~T. 
Resistivity for CeOs$_{4}$Sb$_{12}$ at room temperature (RT) is $\sim500\mu\Omega$ cm which is more than twice as large as $\sim200\mu\Omega$ cm for LaOs$_{4}$Sb$_{12}$. The large residual resistivity ratio (RRR) of $\sim100$ and successful observation of de Haas-van Alphen (dHvA) oscillation for LaOs$_{4}$Sb$_{12}$,~\cite{Sugawara2} could be an indirect evidence of the high quality of CeOs$_{4}$Sb$_{12}$ single crystals grown in the same manner.

\section{RESULTS AND DISCUSSION}
\begin{figure}[h]
\begin{center}\leavevmode
\includegraphics[width=0.9\linewidth]{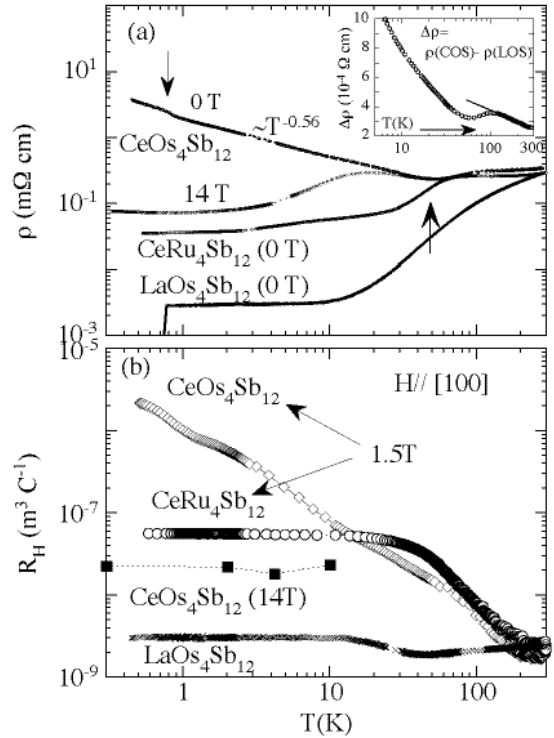}
\caption{Temperature dependence of (a) the electrical resistivity $\rho(T)$ and (b) Hall coefficient $R_{\rm H}(T)$ in CeOs$_4$Sb$_{12}$ along with LaOs$_4$Sb$_{12}$ and CeRu$_4$Sb$_{12}$. The inset of Fig.~\ref{R_RH}~(a) shows the 4$f$ component $\Delta\rho=\rho{\rm (CeOs_{4}Sb_{12})}- \rho{\rm (LaOs_{4}Sb_{12})}$.}
\label{R_RH}
\end{center}
\end{figure}

The temperature dependence of electrical resistivity $\rho(T)$ for CeOs$_{4}$Sb$_{12}$ is compared with those for LaOs$_{4}$Sb$_{12}$ and CeRu$_{4}$Sb$_{12}$ in Fig.~\ref{R_RH}~(a).~\cite{AbeK}
%%%%%%%%%%%%%%%%%%%%%%%%
%%%%%%%%%%%%%%%%%%%%%%%%%
$\rho(T)$ for CeOs$_{4}$Sb$_{12}$ increases approximately as $T^{-1/2}$ over the wide temperature range below $\sim30$~K at 0~T, which qualitatively agrees with the previous report.~\cite{Bauer2} A small but apparent bend in $\rho(T)$ curve is found at around 0.8~K where some phase transition has been reported in the specific heat measurements.~\cite{Namiki}
 
Hedo {\it et al.} reported the resistivity under high pressures,~\cite{Hedo} where they fitted as $\rho(T)\sim {\rm exp}[(T^*/T)^{1/2}]$ and found the reciprocal characteristic temperature $1/T^{*}$ to be proportional to the applied pressure up to 8 GPa. The dependence of $\rho$ on both temperature and pressure has been analyzed on the base of variable range hopping model,~\cite{Efros} though the intrinsic mechanism has not been well described.
 
Recently, Yogi {\it et al.} performed the Sb-nuclear quadrupole resonance (NQR) on this compound and found that the temperature dependence of nuclear-spin-lattice-relaxation-rate $1/T_{\rm 1}$ obeys a relation $1/T_{\rm 1}\sim T^{1/2}$ approximately in the same temperature range (below 25~K).~\cite{Yogi} In the self-consistent renormalization (SCR) theory for the spin fluctuations in itinerant antifferomagnetic (AFM) metals,~\cite{Ishigaki} $1/T_{\rm 1}$ is expected to be proportional to $T^{1/2}$ well above N\'{e}el temperature. However, the contribution from AFM-fluctuation to $\rho$ is expected to decrease with decreasing temperature as $T$ or $T^{3/2}$ depending on the temperature range,~\cite{Moriya} which is inconsistent with $\rho(T)$ in the present measurements. It should be noted that the theories up to now  assume a metal without temperature dependence of carrier densities. The more systematic studies on different properties are necessary to settle the origin of the temperature dependence on $\rho$.
 
The inset of Fig.~\ref{R_RH} (a) shows the 4$f$ component $\Delta\rho=\rho{\rm (CeOs_{4}Sb_{12})}- \rho{\rm (LaOs_{4}Sb_{12})}$, assuming their Fermi surface to be basically the same. Such an assumption might be reasonable at high temperatures, taking into account the closeness of $R_{\rm H}$ near RT.  $\Delta\rho$ increases logarithmically with decreasing temperature down to $\sim100$~K, where it shows a faint maximum. Above 60~K, the transport properties are similar to those of ordinary heavy fermion compounds. After showing a shallow minimum at 60~K, $\Delta\rho$ increases and varies approximately as $T^{-1/2}$ below $\sim30$~K, that is consistent with the fact that the optical conductivity in the low energy range starts to decrease below about 60~K.~\cite{Matsunami}
  
As origins of this resistivity maximum, there are two possibilities depending on the magnitude of Kondo temperature $T_{\rm K}$ compared with the crystalline electric field (CEF) splitting.  In the Ce compounds such as CePd$_3$ with relatively high $T_{\rm K}$ comparable with the CEF excitation $\Delta$, the temperature of the resistivity maximum ($\sim100$~K) roughly corresponds to $T_{\rm K}$.~\cite{Thompson}  In such systems, the magnetic susceptibility $\chi$ exhibits a peak near $T_{\rm K}$, however, $\chi$ in CeOs$_{4}$Sb$_{12}$ monotonously increases with decreasing temperature without any sign of peak structure.~\cite{Bauer2}  On the other hand, in low $T_{\rm K}$ Ce compounds such as CeAl$_{2}$, double peaks in $\rho(T)$ have been observed as a function of temperature,~\cite{Onuki} and are related with the two Kondo temperatures; $T_{\rm K}^h$ at high temperatures associated with all the CEF split levels and $T_{\rm K}^\ell$ for low temperatures resulting from only the CEF ground state for Ce-heavy fermion compounds. According to Hanzawa {\it et al.},~\cite{Hanzawa} $T_{\rm K}^h$ and $T_{\rm K}^\ell$ are related as
%%%%%%%%%%%%%%%%%%%%%%%%%
\begin{equation}
T_{\rm K}^h=(T_{\rm K}^\ell\Delta_1\Delta_2)^{1/3}
\label{eq1}
\end{equation}
%%%%%%%%%%%%%%%%%%%%%%%%%
where $\Delta_1$ and $\Delta_2$ are the CEF-splitting between the ground state and the first and the second excited states, respectively.  Putting $\Delta=\Delta_1=\Delta_2= 327$~K (between $\Gamma_{\rm 7}$ ground state and $\Gamma_{\rm 8}$ excited state estimated from the temperature dependence of magnetic susceptibility~\cite{Bauer2}) and $T_{\rm K}^h=100$~K at the resistivity maximum, $T_{\rm K}^\ell$ is estimated as $\sim10$~K, leading to the specific heat coefficient $\gamma\sim1000$~mJ/K$^2$mol.~\cite{Onuki2}  The experimental value of specific heat coefficient $\gamma\sim180$~mJ/K$^2$mol,~\cite{Namiki} is rather close to that in the high $T_{\rm K}$ scenario of $\gamma\sim100$~mJ/K$^2$mol, which contradicts with the experimental result. However, it should be noted that the peak temperature $\sim18$~K under 14~T is close to the estimated $T_{\rm K}^\ell$ and the magnetic contribution to the resistivity at 0 T shown in the inset of Fig.~2(a) follows $-lnT$ dependence only within the narrow temperature range $10-30$~K.

Both the dc and the optical conductivities indicate the decreasing carrier number below $\sim60$~K, which suggests the model to explain the low temperature properties of this material must have the temperature dependent carrier number and electronic density of states.
$\rho(T)$ above $\sim60$~K can be understood as of the ordinary Ce Kondo compound with a peak at $\sim100$~K corresponding to $T_{\rm K}$ or $T_{\rm K}^h$. At lower temperatures, the increases in $\rho$ and $R_{\rm H}$ indicate the reduction of density of states at $E_{\rm F}$. However, the approximate $T^{-1/2}$ dependence of $\rho$ in such a wide temperature range (of about two orders of magnitude) down to 0.6~K rules out the simple semiconducting state as an origin.

Such a temperature dependence of $\rho$ could be ascribed to the temperature dependence of carrier density $n$ and scattering lifetime of electrons ${\bf \tau}$, even if we assume the simplest single FS model. By combining $R_{\rm H}$, where only the temperature dependence of $n$ plays a role, we might be able to separate the two contributions. Figure~\ref{R_RH}~(b) shows the temperature dependence of $R_{\rm H}$ for CeOs$_{4}$Sb$_{12}$ along with those for the reference compounds LaOs$_{4}$Sb$_{12}$ and CeRu$_{4}$Sb$_{12}$.  At high temperatures, $R_{\rm H}$ is positive for all the three compounds and the magnitude is not much different. The estimated carrier density is between $1.0-1.4$~holes/f.u. at RT. The $T$-dependence of $R_{\rm H}$ for LaOs$_{4}$Sb$_{12}$ is not so large, but have a broad minimum near $\sim40$~K. The decrease down to 40~K is ascribed to the change in the main scattering centers from the isotropic phonon-scattering with large wave vectors (${\bf q}$) to the anisotropic phonon-scattering with smaller ${\bf q}$, and the increase below $\sim40$~K reflects the recovery to the isotropic scattering by impurities.~\cite{Tsuji,Sato3}

\begin{figure}[h]
\begin{center}\leavevmode
\includegraphics[width=0.8\linewidth]{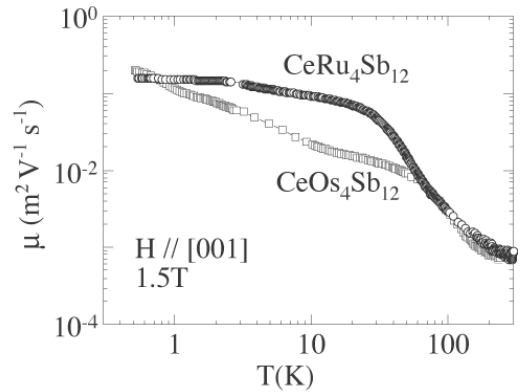}
\caption{Temperature dependences of Hall mobility $\mu$ in CeOs$_{4}$Sb$_{12}$ and CeRu$_{4}$Sb$_{12}$.}
\label{mu}
\end{center}
\end{figure}

\begin{figure}[h]
\begin{center}\leavevmode
\includegraphics[width=0.9\linewidth]{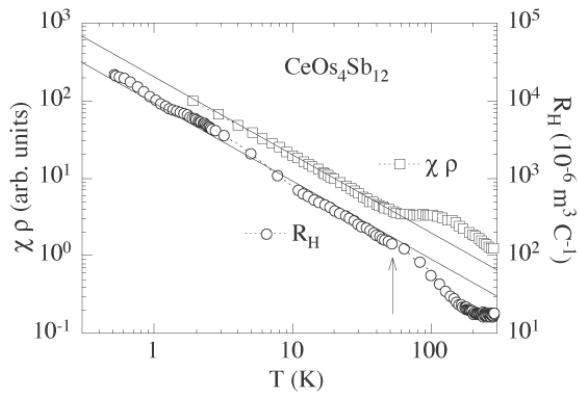}
\caption{Temperature dependences of $\chi\rho$ in CeOs$_{4}$Sb$_{12}$: analysis of the anomalous Hall effect (skew scattering).}
\label{chi_rho}
\end{center}
\end{figure}

Near RT, $R_{\rm H}$ for both CeOs$_{4}$Sb$_{12}$ and CeRu$_{4}$Sb$_{12}$ increases drastically with decreasing temperatures. Such an increase in $R_{\rm H}$ have been observed in many Ce-based Kondo compounds related with the Kondo-like increase in resistivity,~\cite{Onuki3} however, the magnitude is unusually larger than that was reported previously; usually $R_{\rm H}$ shows a peak of the magnitude less than $10^{-8}$m$^3$/C near $T_{\rm K}$. For CeRu$_{4}$Sb$_{12}$, $R_{\rm H}$ tends to saturate to $5.7\times10^{-8}$m$^3$/C below $\sim50$~K, which is consistent with the temperature dependence of the electronic density of states at $E_{\rm F}$ in the optical measurement.~\cite{Kanai} Interestingly, $R_{\rm H}$ in CeOs$_{4}$Sb$_{12}$ shows approximately $T^{-1}$ dependence in the same temperature range where $\rho$ varies approximately as $T^{-1/2}$. This fact automatically rules out the simple assignment of the origin of the $T^{-1/2}$ dependence in resistivity to the $T$-dependence of carrier density.  In the case of Kondo insulator CeNiSn, the inconsistent temperature dependence between $\rho$ and $R_{\rm H}$ (the decrease in $\rho$ contradicts with the decreasing carrier number estimated from $R_{\rm H}$ with decreasing temperature) has been ascribed to the increase in the carrier mobility with decreasing temperature.~\cite{Takabatake} The temperature dependences of Hall mobility $\mu=R_{\rm H}/\rho$ in CeOs$_{4}$Sb$_{12}$ and CeRu$_{4}$Sb$_{12}$ are compared in Fig~\ref{mu}.
%%%%%%%%%%%%%%%%%%%%%%%%%%%%%%
%%%%%%%%%%%%%%%%%%%%%%%%%%%%%%
Both the magnitude and the temperature dependence of $\mu$ for the two compounds are very close above $\sim60$~K, where no drastic change in the electronic density of states at $E_{\rm F}$ has been reported.~\cite{Matsunami,Kanai} Thus, the initial rise in $\mu$ with decreasing temperature in Fig.~\ref{mu} could not be ascribed to the increase in the carrier mobility within the normal Hall effect contribution, but is more naturally ascribed to the anomalous Hall effect (the skew scattering). In addition, for CeOs$_{4}$Sb$_{12}$, taking into account $T^{1/2}$ dependence of $1/T_1$ associated with the spin fluctuation near AFM critical point, the anomalous Hall component is expected to dominate also at lower temperatures.  In many Kondo compounds, the $T$-dependence of Hall coefficient roughly follows Eq.~(\ref{eq2}).~\cite{Fert}
%%%%%%%%%%%%%%%%%%%%%%%%%
\begin{equation}
R_{\rm H}(T)=R_{\rm H0}+R_{\rm S}(T)\chi(T)
\label{eq2}
\end{equation}
%%%%%%%%%%%%%%%%%%%%%%%%%
where $R_{\rm S}(T)$ is a function of the magnetic part of electrical resistivity. $R_{\rm S}(T)$ in Kondo materials has been well described by the skew component under selected conditions.  For rough estimation, we have calculated the skew component based on the simplest assumption of $R_{\rm S}(T)=\alpha\rho(T)$; the coefficient $\alpha$ takes different values above and below $T_{\rm K}$ depending on the phase shift associated with the scattering channels,~\cite{Fert}  $\rho(T)$ is the electrical resistivity under 1.5 T simultaneously measured with Hall coefficient. The result is compared with the experimental one in Fig.~\ref{chi_rho}, where the main characteristic of $R_{\rm H}(T)$; the almost $T^{-1}$ dependence on temperature, is roughly reproduced, taking into account the oversimplification of the model.
%%%%%%%%%%%%%%%%%%%%%%%%%%%%%%
%%%%%%%%%%%%%%%%%%%%%%%%%%%%%%
The difference in curvature above $\sim50$~K may be ascribed to the change in sign of $\alpha$ above and below $T_{\rm K}$.

\begin{figure}[h]
\begin{center}\leavevmode
\includegraphics[width=0.9\linewidth]{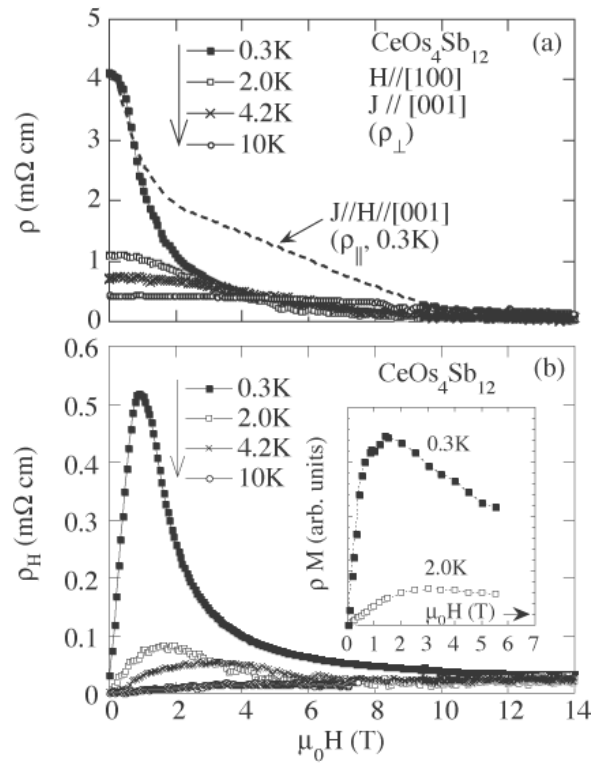}
\caption{Field dependences of (a) the electrical resistivity $\rho(H)$ and (b) Hall resistivity $\rho_{\rm H}(H)$ in CeOs$_{4}$Sb$_{12}$.}
\label{r_H}
\end{center}
\end{figure}

To further understand the unusual temperature dependence of $\rho(T)$ and $R_{\rm H}(T)$ in Figs.~\ref{R_RH}~(a) and \ref{R_RH}~(b) at low temperatures, the resistivity and Hall resistivity at selected temperatures have been measured as a function of magnetic field in Figs.~\ref{r_H}~(a) and \ref{r_H}~(b), respectively.
%%%%%%%%%%%%%%%%%%%%%%%%%%%%%%
%%%%%%%%%%%%%%%%%%%%%%%%%%%%%%
$\rho(H)$ is dramatically suppressed; especially at 0.3 K, it is reduced to $\sim1/70$ under 14~T. $\rho_{\rm H}$ exhibits a peak near 0.9~T above which it also shows a drastic reduction. In the form $R_{\rm H}^*= \rho_{\rm H}/\mu_0H$, the reduction factor $\sim1/350$ is about five times larger than that in $\rho$, which is unexplainable only by a carrier density change in the simplest single carrier model. This drastic field effect is in sharp contrast with the pressure effect on $\rho$ reported to be minor in CeOs$_{4}$Sb$_{12}$.~\cite{Hedo}
 
These field dependences might be ascribed to the combined effect of changes in carrier number and in carrier scattering intensity. The former alone in the simplest single carrier model is unable to explain both $\rho(T)$ and $R_{\rm H}(T)$ as was already mentioned related with the mobility in Fig.~\ref{mu}. 
In magnetic systems, the anomalous Hall component sometime predominates over the normal one; i.e., the Hall resistivity composed of the normal component proportional to magnetic field $H$ and anomalous one proportional to magnetization $M$ as
%%%%%%%%%%%%%%%%%%%%%%%%%
\begin{equation}
\rho_{\rm H}(H)=R_{\rm H0}H+R_{\rm S}M(H)
\label{eq3}
\end{equation}
%%%%%%%%%%%%%%%%%%%%%%%%%
Using $\rho(H)$ in Fig.~\ref{r_H}~(a) and $M(H)$ measured by a SQUID magnetometer, the second term in Eq.~(\ref{eq3}) is plotted in the inset of Fig.~\ref{r_H}~(b).~\cite{M}  The peak structure can be reproduced, however, the agreement of both the position and the width are not satisfactory. It must be noted that the theory on the anomalous Hall effect assumes ordinary metallic Kondo systems with a basically constant carrier concentration. To make quantitative comparison, a model taking into account the temperature dependence of carrier number is necessary.

\begin{figure}[h]
\begin{center}\leavevmode
\includegraphics[width=0.9\linewidth]{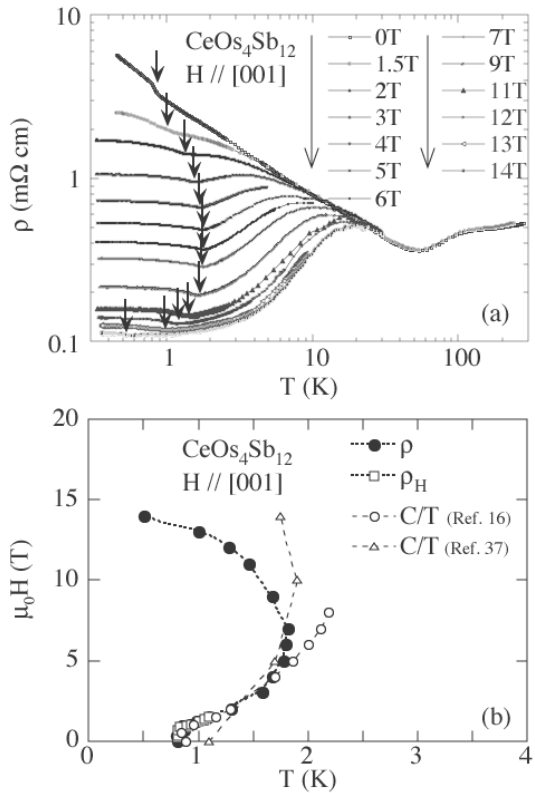}
\caption{(a) Temperature dependence of the electrical resistivity under selected magnetic fields and (b) $H-T$ phase diagram in CeOs$_{4}$Sb$_{12}$. The higher field triangle point in the phase diagram was estimated from a faint peak in the specific heat in Ref.~37.}
\label{PhaseDiagram}
\end{center}
\end{figure}

To understand another characteristic feature in Fig.~\ref{R_RH}, a small but clear anomaly at 0.8~K where some phase transition was found in the specific heat measurements,~\cite{Bauer,Namiki} we have measured the temperature dependence of $\rho$ as shown in Fig.~\ref{PhaseDiagram}~(a) along with $\rho_{\rm H}$ (not shown) under selected magnetic fields.
%%%%%%%%%%%%%%%%%%%%%%%%%%%%%%
%%%%%%%%%%%%%%%%%%%%%%%%%%%%%%
The position of the anomaly shifts with magnetic field, which is plotted in Fig.~\ref{PhaseDiagram}~(b) as a $H-T$ phase diagram where the anomalies found in the specific heat measurements are also plotted.~\cite{Namiki,Rotundu}  Above about 4~T, the phase boundary determined by the transport measurements deviates from that by the specific heat measurement, which might be due to the small missalignment of crystalline direction to the magnetic field.

Recently, Rotundu and Andraka also determined the $H-T$ phase diagram for $H\|[100]$ up to 10~T based on the specific heat measurements.~\cite{Rotundu} Their data points are in between those in the present experiment and those in the specific heat measurement by Namiki {\it et al}. below 10~T above which the anomaly is almost invisible in $C(T)$. On the other hand, $\rho(T)$ in the present measurements exhibits an evident anomaly up to 14~T, though the magnitude becomes quite small.

The feature of phase diagram is reminiscent of the antiferro-quadrupole (AFQ) transition observed in CeB$_6$.~\cite{Fujita} However, as already pointed out from the $C(T)$ measurements,~\cite{Bauer,Namiki} the electronic part of the entropy release ($0.05{\rm R}ln2$ at 0 T and $0.06{\rm R}ln2$ at 4 T) below the transition temperature is too small to be attributed to localized $f$-electron contributions. Itinerant nature of $4f$ electrons is also suggested by Bauer {\it et al.} in relation to the Wilson ratio $R_{\rm W}=(\pi^2 k_{\rm B}^2/3\mu_{\rm eff}^2)(\chi_0/\gamma)$ which is of the order of unity for CeOs$_4$Sb$_{12}$.~\cite{Bauer2} Here $\mu_{\rm eff}$ is the effective magnetic moment, $\chi_0$ is the Pauli susceptibility, and $\gamma$ is the Sommerfeld coefficient. Moreover, the possibility of an AFQ transition in cubic system requires that the CEF ground state is $\Gamma_8$ quartet, however, from the magnetic susceptibility measurements,~\cite{Bauer2} the  $\Gamma_7$ ground state is suggested for CeOs$_4$Sb$_{12}$ inconsistent with the AFQ scenario. Taking into account the minor change in $\rho$ and $R_{\rm H}$ across the transition along with the anomalous Hall contribution reflecting the conduction electron scattering by magnetic instability, the transition at 0.8~K may be ascribed to the spin-density-wave (SDW) order.  Yogi {\it et al}. have found a clear anomaly at ${\sim0.9}$~K in a recent Sb-NQR experiment,~\cite{Yogi} which was ascribed to the onset of SDW order.

Rotundu and Andraka pointed out that their finding of a sizable decrease of Sommerfeld coefficient above 5~T is in sharp contrast with the Kondo insulators,~\cite{Rotundu} such as CeNiSn; where magnetic fields increase the Sommerfeld coefficient by destroying the $c-f$ hybridization and closing the energy gap.~\cite{Takabatake2} The quenching of spin fluctuations by magnetic fields is a possible explanation for the apparent decrease of Sommerfeld coefficient,~\cite{Ikeda} which also indirectly suggests the origin of phase transition to be the SDW order. However, the relation between the SDW order and the AFQ-like phase diagram is still unclear. Ce(Ru$_{1-x}$Rh$_x$)$_2$Si$_2$ ($x=0.05-0.25$) is also reported to exhibit a SDW order below 6 K.~\cite{Sekine2} However, there exists a clear difference between the two compounds; in Ce(Ru$_{1-x}$Rh$_x$)$_2$Si$_2$, the change in carrier number may be minor, even if it exits, since it has a metallic ground state and the $H-T$ phase diagram exhibits an ordinary AFM-like one. Therefore, one can infer that the change in carrier density might play some role in the unusual $H-T$ phase diagram of CeOs$_4$Sb$_{12}$. To elucidate the origin of the phase transition and to establish the phase diagram including anisotropy, more intense studies on other physical properties are necessary.

Finally, we discuss the comparison of the transverse and longitudinal magnetoresistance as shown in Fig.~\ref{r_H}~(a). $\rho$ is larger in the longitudinal ($\rho_{\rm \|}$) than in the transverse geometry ($\rho_{\rm \perp}$) above $\sim1$~T. If only the Lorenz magnetoresistance plays a role, $\rho_{\rm \perp}$ is never smaller than $\rho_{\rm \|}$. If one assumes the conventional form of exchange interaction $V=-J{\bf s\cdot S}$ between a conduction electron with spin ${\bf s}$ and a magnetic impurity with spin ${\bf S}$, the magnetoresistance is isotropic; independent of the relative orientation of the current and magnetic field directions. Fert reported experimental results of anisotropic magnetoresistance due to heavy rare earth ions in Au,~\cite{Fert2} which is explained by taking into account the quadrupolar interaction. He has also calculated the anisotropic magnetoresistance for Ce impurities in La,~\cite{Fert3} which qualitatively explains the experimental result.  At this stage, we cannot say any conclusive remark on the anisotropic magnetoresistance in Fig.~\ref{r_H}~(a), however, it should be noted that some contribution from the orbital angular moment is necessary to explain such a anisotropic magnetoresistance. That might be also related with the unusual $H-T$ phase diagram, in which the strong correlation between 4$f$- and conduction electrons is expected.  

In summary, We have found an unusual temperature dependence of both electrical resistivity and Hall coefficient below $\sim30$~K in CeOs$_4$Sb$_{12}$, which may be ascribed to the combined effect of carrier density decreasing and spin fluctuations. The anomalous low temperature state with large electrical resistivity is considerably suppressed by the magnetic fields, suggesting a drastic change of electronic structure and  suppression of spin fluctuations by the magnetic field.

\begin{acknowledgments}
The authors are grateful to thank Professors. H. Harima, M. Kohgi, K. Iwasa, O. Sakai, Y. \={O}nuki, K. Miyake, Y. Kitaoka, Drs. M. Yogi, and T. Takimoto for helpful discussions. They also thank Mr. A. Kuramochi for his help in the transport measurements. This work was supported by a Grant-in-Aid for Scientific Research Priority Area "Skutterudite" (No.15072206) of the Ministry of Education, Culture, Sports, Science and Technology, Japan.
\end{acknowledgments}

\newpage

%%%%%%%%%%%%%%%%%%%%
%\begin{figure}[]

%\caption{Energy gap  $\Delta E_{\rm g}$ $vs$ lattice constant in CeTr$_{4}$Pn$_{12}$.~~~~~~~~~~~~~~~~~~~~~~~~~~~~~~~~~~~~~~~~~~~~~~~~~~~~~~~~~~~~}
%\label{Eg}

%\caption{Temperature dependence of (a) the electrical resistivity $\rho(T)$ and (b) Hall coefficient $R_{\rm H}(T)$ in CeOs$_4$Sb$_{12}$ along with LaOs$_4$Sb$_{12}$ and CeRu$_4$Sb$_{12}$. The inset of Fig.~\ref{R_RH}~(a) shows the 4$f$ component $\Delta\rho=\rho{\rm (CeOs_{4}Sb_{12})}- \rho{\rm (LaOs_{4}Sb_{12})}$.}
%\label{R_RH}

%\caption{Temperature dependences of Hall mobility $\mu$ in CeOs$_{4}$Sb$_{12}$ and CeRu$_{4}$Sb$_{12}$.~~~~~~~~~~~~~~~~~~~~~~~}
%\label{mu}

%\caption{Temperature dependences of $\chi\rho$ in CeOs$_{4}$Sb$_{12}$: analysis of the anomalous Hall effect (skew scattering).}
%\label{chi_rho}

%\caption{Field dependences of (a) the electrical resistivity $\rho(H)$ and (b) Hall resistivity $\rho_{\rm H}(H)$ in CeOs$_{4}$Sb$_{12}$.}
%\label{r_H}

%\caption{(a) Temperature dependence of the electrical resistivity under selected magnetic fields and (b) $H-T$ phase diagram in CeOs$_{4}$Sb$_{12}$. The higher field triangle point in the phase diagram was estimated from a faint peak in the specific heat in Ref.~37.}
%\label{PhaseDiagram}

%\end{figure}
%%%%%%%%%%%%%%%%%%%%%%%%%%%%%%


\begin{thebibliography}{99}
\bibitem{M}In the analysis of $\rho M$ at 0.3 K, we used $M(H)$ at 2~K since our SQUID magnetometer works only down to $\sim2$~K. No essential change of $M(H)$ is expected at least below 3~K.
\end{thebibliography}
\end{document}